\documentclass[twocolumn,amssymb, nobibnotes, aps, amsmath, amssymb, superscriptaddress,floatfix,]{revtex4-1}

\usepackage[utf8]{inputenc}
\usepackage[T1]{fontenc}
\usepackage{graphicx}
\usepackage{dcolumn}
\usepackage{bm}
\usepackage{newtxtext, newtxmath}  
\DeclareGraphicsExtensions{.pdf,.jpeg,.jpg, .png, .eps, .tiff}
\usepackage{epstopdf}
\usepackage{lipsum}
\usepackage{tabularx}
\usepackage{booktabs}
\usepackage{array}
\usepackage{natbib}
\usepackage[dvipsnames]{xcolor}
\usepackage{grffile}
\usepackage{boxhandler}
\usepackage{float}
\usepackage[normalem]{ulem}
\usepackage{hyperref}
\usepackage{wasysym}
\usepackage{soul}
\usepackage{siunitx}

\newcommand{\pea}{\text{Pe}_\text{a}}
\newcommand{\pes}{\text{Pe}_\text{s}}
\newcommand{\gp}{G'}
\newcommand{\gpp}{G''}
\newcommand{\peclet}{P\'{e}clet }

\newcommand{\fig}[1]{\textbf{Fig.~\ref{#1}}}
\newcommand{\FIG}[1]{\textbf{Figure~\ref{#1}}}
\newcommand{\eq}[1]{\textbf{Eq.~\ref{#1}}}

\newcommand{\ue}{School of Physics and Astronomy, The University of Edinburgh, Peter Guthrie Tait Road, Edinburgh, EH9 3FD, United Kingdom}
\newcommand{\uc}{Niels Bohr Institute, University of Copenhagen, Blegdamsvej 17, Copenhagen, Denmark.}
\newcommand{\ucmDE}{Departamento de Estructura de la Materia, Física Térmica y Electrónica, Universidad Complutense de Madrid, Madrid, Spain.}
\newcommand{\ucmDQF}{Departamento de Química Física, Facultad de Química, Universidad Complutensede Madrid, Madrid, Spain.}

\newcommand{\cu}{Department of Chemistry, Columbia University, New York, NY, 10027, USA.}

\definecolor{pumpkin}{rgb}{1.0,0.4,0.0}
\definecolor{mygreen}{rgb}{0.0,0.55,0.3}
\definecolor{strawberry}{rgb}{1.0,0.0,0.5}
\definecolor{midnight}{rgb}{0.003921569,0.098039216,0.576470588}
\definecolor{saphire}{rgb}{0.0,0.196,0.372549}
\definecolor{crimson}{rgb}{0.75686,0,0.262745}
\definecolor{capri}{rgb}{0.0,0.768627,0.8745098}

\newcommand{\correctMath}[2]{{\textcolor{NavyBlue}{#2}}}

\begin{document}

\title{MIPS is Maxwell-like fluid with an extended and non-monotonic crossover}

\author{Jos\'{e} Mart\'{i}n-Roca}
\affiliation{\ucmDE}\affiliation{\ucmDQF}
\author{Kristian Thijssen}
\affiliation{\uc}
\author{Tyler Shendruk}
\email{t.shendruk@ed.ac.uk}
\affiliation{\ue}
\author{Angelo Cacciuto}
\email{ac2822@columbia.edu}
\affiliation{\cu}
\author{Chantal Valeriani}
\email{cvaleriani@ucm.es}
\affiliation{\ucmDE}

\newcommand{\supprefs}{\cite{lees1972computer,allen2017computer,mandal2021shear,villarroel2021critical,winkler2015virial,hatwalne2004rheology,petrelli2020effective,van1989hard}}

\begin{abstract}
    Understanding the mechanical properties of active suspensions is crucial for their potential applications in materials engineering. Among active phenomena with no analogue in equilibrium systems, motility-induced phase separation (MIPS) in active colloidal suspensions is one of the most extensively studied. However, the mechanical properties of MIPS remain poorly understood.  This study investigates the rheology of a suspension of active colloidal particles under constant and oscillatory shear. Systems consisting of pseudo-hard active Brownian particles exhibiting co-existence of dense and dilute phases behave as a viscoelastic {Maxwell-like} fluid at low and high frequencies, displaying exclusively shear thinning across a wide range of densities and activities. {Remarkably, the crossover frequency between the storage and loss moduli is non-monotonic, increasing with activity before the MIPS transition but decreasing with activity after the transition, in contrast to a passive analog system of attractive particles in liquid-gas phase coexistence}, revealing the subtleties of how  active forces and intrinsically out-of-equilibrium phases affect the mechanical properties.
\end{abstract}

\maketitle

Active materials, composed of self-propelled units that consume energy to generate motion, display a variety of fascinating properties that deviate from those of systems at equilibrium~\cite{ramaswamy2017active,bechinger2016active}, including large correlated motion~\cite{baconnier2025self,mallory2018active}, anomalous transport~\cite{williams2022confinement,liao2019mechanism,martin2025carnivorous,granek2022anomalous,koumakis2014directed}, and emergent self-organization~\cite{ramaswamy2010mechanics, granek2022anomalous, rana2019tuning, peng2024self}.  
Motility-Induced Phase Separation (MIPS), where a solution of motile particles separates into co-existing dense and dilute regions without attractive interactions, is a textbook example of behavior exclusive to active systems~\cite{cates2015motility} that has been the subject of intense scrutiny~\cite{torres2024motility,martin2021characterization,stenhammar2014phase,kolb2020active,cates2013active,solon2015pressure}. 
This is because active Brownian particles (ABP) represent the simplest active model  with a well-defined interplay between thermal, dispersion and active forces~\cite{marchetti2016minimal}. 
The intrinsically out-of-equilibrium nature of these self-organized structures  implies that traditional approaches and even definitions that are workhorses in equilibrium statistical mechanics must be applied with care. A striking example is the use of conflicting definitions of surface tension to study MIPS interfaces~\cite{bialke2015,patch2018,omar2020,siebert2018,theurkauff2015,fily2014,takatori2014,mallory2021,chacon2022intrinsic,hermann2019}. 

Despite standing as a fundamental state of active matter, the mechanical properties of the MIPS phase remain poorly understood, especially its response to external driving forces.  
Much of the work on externally driven ABPs has focused on  dense or crystalline suspensions in which activity fluidizes otherwise "jammed" systems~\cite{ongenae2021activity}, inducing either shear thinning~\cite{wiese2023fluid} or thickening~\cite{bayram2023motility} depending on particle softness. 
In contrast, the rheological properties of dilute ABPs, where activity drives large-scale density variations through MIPS~\cite{cates2015motility}, is largely unexplored. 

Crucially, while MIPS is often described as a liquid-gas phase separation due to its mathematical analogy with equilibrium systems~\cite{hecht2024motility}, a significant degree of local crystalline order is detectable within the dense phase~\cite{cates2015motility}. 
The translational symmetry breaking suggests that the dense phase is more akin to a solid crystal than a liquid droplet. 
This apparent paradox raises key questions about the rheological nature of such systems, since broken symmetries generally result in elasticity --- at least in passive materials. 
Traditionally, the rheological properties of passive systems arise from a thermal relaxation in response to external driving.  However, active materials possess intrinsically out-of-equilibrium pathways by which emergent phenomena and novel behaviors can arise in response to driving forces, such as super-fluid like behavior~\cite{takatori2017superfluid}, the emergence of higher-order defects~\cite{rivas2020}, non-linear Darcy's law~\cite{mackay2020,keogh2024} and negative drag~\cite{Foffano2012}.

While active matter holds promise for developing exciting materials with unusual mechanical properties, understanding how these materials respond to external driving forces is key to comprehending and designing their mechanical behavior. 
Studying the rheology of the MIPS phase is a first fundamental step in this direction.
In this work, we study the rheological behavior of a suspension of pseudo-hard ABPs across the MIPS transition using numerical simulations.
{ Specifically, we measure the rheology of this system in its homogeneous fluid state, and in the phase separated (MIPS) state, which accounts for the simultaneous presence of the low density gas and the high density condensate.  }
We consider both oscillatory and steady shear, and  show  the rheological response is akin to a {Maxwell-like} fluid at low and high frequencies, but has an extended crossover regime at intermediate frequencies. 
In contrast with the monotonic behavior observed in the  storage/loss moduli crossover frequency for passive attractive systems, we discover that the crossover point is non-monotonic with activity since the timescale switches from one associated with thermal processes to one associated with active motion.

\begin{figure}[tb]
    \centering
    \includegraphics[width=1\columnwidth]{Figure_1.png}
    \caption{ Storage moduli $\gp$ (closed circles) and loss moduli $\gpp$ (open squares) with oscillation amplitude $A_x/L_y=5\%$ (SAOS regime) for $\rho=0.5$ and different activities $\pea  = 0$ (blue), $42$ (green) and $120$ (red).
    Dashed and straight lines are visual guides to show the expected {M}axwell scaling of $\omega^1$ and $\omega^2$.
    } 
    \label{fig:fig1}
\end{figure}


We perform Brownian dynamics simulations of ABP of diameter $d$ in two dimensions with translation diffusion $D_\text{T}$ and rotation diffusion $D_\text{R}=3D_\text{T}/d^2$.

The self-propulsion speed can be imagined as due to an active force of strength $F_a=\zeta \, v_0$. The excluded volume interaction between the particles is enforced via the pairwise pseudo-hard sphere potential~\cite{jover2012pseudo}. We resolve the equations of motion using LAMMPS~\cite{LAMMPS} (see SI). The number density is $\rho=0.5$, unless otherwise stated. 

{\color{black} We impose shear through Lees–Edwards boundary conditions, which ensure that shear strain is transmitted homogeneously throughout the system.} When a strain $\gamma$ with strain rate $\dot{\gamma}$ is applied to the active system (see SI), the dynamics of individual particles are governed by two dimensionless numbers $\pes$ and $\pea$~\cite{bayram2023motility}. The shear \peclet number, $\pes = \frac{\dot{\gamma} \, d^2}{D_\text{T}}$,
provides information about the relevance of the imposed shear rate with respect to thermal fluctuations, while the Active \peclet number, $\pea =\frac{3v_0}{D_\text{R} d}$,
measures the ratio between the orientation decorrelation time $\tau_R = 1/D_\text{R}$ and self-propulsion time $\tau_\text{a}=d/v_0$.

If the amplitude of the applied deformation is sufficiently small, the system responds linearly with stress (see SI) proportional to strain, $\gamma(t) = ( A_x / L_y ) \, \sin(\omega \, t)$ and strain rate, $\dot{\gamma}(t)= ( A_x \omega / L_y ) \, \cos(\omega \, t)$, where $\omega$ is the frequency of the oscillatory shear.  
In this regime, only the first-order natural modes of the system are excited ~\cite{barnes1989introduction, daalkhaijav2018rheological}, and the stress response is $\sigma_{xy}(t) = \gp \gamma(t) + \gpp \dot{\gamma}(t) / \omega$, where $\gp$ and $\gpp$ are the storage and loss moduli, respectively.
To ensure a linear response, a fixed-frequency amplitude sweep is performed across a broad range of frequencies (SI~\cite{SI}).

Figure~\ref{fig:fig1} shows the interplay between the elastic and viscous responses for different values of $\pea$,  below (blue), above (red) and at (green) the MIPS transition (see SI~\cite{SI}). 
Interestingly, regardless of $\pea$, the system shows a liquid-like response $(\gpp>\gp)$ at low frequencies (small $\pes$). 
While this is expected for passive systems characterized by a single low density fluid phase (blue data), the behavior above the MIPS transition point (red data) indicates that the elastic response is still compatible with that of a liquid rather than  a solid, despite the high degree of crystalline order observed within the fully formed condensed phase (\fig{fig:fig2}; { top} inset). 
This is unexpected and suggests that  particle rearrangement occurs within the active condensed clusters under the action of low-frequency shear forces.  In all cases, higher values of Pe$_a$ imply higher values of $G'$ and $G''$, since dissipative processes must enhance $G''$ away from equilibrium and increased $G'$ arises from jammed-particle contact in MIPS. As the frequency increases, both moduli increase with $\gpp\sim \omega$ and $\gp\sim \omega^2$.
Interestingly, these exponents are precisely what one expects for Maxwell-like fluids in the low frequency limit~\cite{grimm2011brownian}. 

\begin{figure}[tb]
    \centering
    \includegraphics[width=0.92\columnwidth]{Figure_2.pdf}
    \caption{Dimensionless crossover frequency {\color{black}$\pes^\dagger= \frac{A_xd^2}{L_y D_t} \omega_c$} for storage $\gp$ and loss $\gpp$ moduli shown in \fig{fig:fig1} as a function of dimensionless activity $\pea$ at density $\rho =0.5$ for oscillation amplitude $A_x/L_y=5\%$ (SAOS regime). Shaded area represents the MIPS phase separation boundary  (see SI~\cite{SI}).   \textbf{Insets:} Snapshots corresponding to the colored points and semi-log plot.  {\color{black} The dashed curve fits to $\omega_c\sim a_\omega+b_\omega\, \pea^2$ for $\pea<5$ and continuous curve fits to $\omega_c \sim a_\omega'+b'_\omega\, \pea$ for $5<\pea<30$ (see SI for more details). } } 
    \label{fig:fig2}
\end{figure}

{ 
At high frequencies independent of $\pea$, the storage modulus $\gp$ approaches a plateau  $G'(\dot{\gamma}\to \infty)\equiv G'_\infty$ (\fig{fig:fig1}). This occurs in both the passive attractive and active systems. However, in the passive attractive case $G'_\infty$ is similar at all temperatures studied (see SI~\cite{SI}), while in the active case it varies as a function of Pe$_a$}. Simultaneously, the loss modulus $\gpp$ is drastically reduced as the oscillation period becomes much shorter than the typical relaxation time, causing the system to behave as a purely elastic solid. In this regime, system deformation and the displacement of the particles are synchronized, resulting in  $\gpp \to 0$, as expected for a Maxwell fluid. When comparing the purely repulsive active system in the MIPS phase to the passive attractive system in liquid-gas coexistence (see SI~\cite{SI}), $G''$ shares similarities both at low and high frequencies, such as   the broadness of the $G''$ curve corresponding to multiple relaxation times.

\begin{figure}[tb]
    \centering
    \includegraphics[width=1\columnwidth]{Figure_3.png}
    \caption{Effective viscosity $\eta$ as a function of dimensionless frequency $\pes$ with oscillation amplitude $A_x/L_y=5\%$ (SAOS regime) for density $\rho=0.5$ and different activities: $\pea =0$ (blue), $42$ (green) and $120$ (red). Viscosity computed from \fig{fig:fig1} using \eq{eq:visc}. 
    \textbf{Inset:} Zero-shear limit $\eta_0=\lim_{\pes\to0}\eta$ as a function of  activity. 
    Colored points are the same as in \fig{fig:fig2}}
    \label{fig:fig3}
\end{figure}

Since $\gp$ rises faster than $\gpp$ at low frequencies, the curves have a crossover frequency, {\color{black} $\omega_c$}, which establishes the dominant relaxation time of the material {\color{black}($\sim 1/\omega_c$)}, and the onset value above which the system exhibits a solid-like elastic response to shear. {\color{black}  At zero activity, the crossover  $\pes^\dagger=\frac{A_xd^2}{L_y D_t} \omega_c \approx 30$ is consistent with structural relaxation in the dense suspension, and is well  described by the Maxwell relation, $\omega_c = \tfrac{G'_\infty}{\eta_0}$~\cite{van1989hard,mason1995optical,khrapak2024stokes}.} As activity is introduced, $\omega_c$ increases  with $\pea$ until it reaches a maximum right before the MIPS transition. At the transition point, a sharp drop in $\omega_c$ occurs. The crossover time is non-monotonic with activity, which contrasts with a suspension of attractive passive particles in liquid-gas phase coexistence state for which the crossover time is monotonic with temperature (see SI~\cite{SI}).

{\color{black} Activity takes two distinct roles. At lower activities, the activity drives faster movement, but at higher activity it drives local crowding through cluster formation. These considerations explain why MIPS has a  non-monotonic crossover point:}  At low activities in the gas-like phase, the active time scale $\tau_\text{a}$ that dominates the response and so the crossover frequency grows {\color{black}quadratic} with $\pea \sim v_0$ {\color{black} (see SI \cite{SI})}.  At the MIPS transition, the crossover point drops sharply because of sudden phase separation. {\color{black} Above the MIPS transition, the crossover is controlled by rotational relaxation of the clusters. Using $D_r = 3D_t/d^2$, the observed plateau $\pes^\dagger=\frac{A_xd^2}{L_y D_t} \omega_c \approx 5$ corresponds to a rotational Weissenberg number $Wi_R = \dot\gamma_c/D_R \approx 1.7$, i.e. of order unity. This reflects the expected criterion that the crossover occurs when the shear rate matches the intrinsic rotational relaxation rate of the clusters.} The crossover is independent of activity because the system is dominated by large clusters and the relaxation time associated with the clusters is $\tau_\text{r}$, independent of activity. 

While the low and high frequency limits are dominated by single timescales, the response at intermediate frequencies around the crossover $\pes^\dagger$ is more complex, exhibiting an extended regime compared to Maxwell fluids (\fig{fig:fig1}), potentially due to the existence of multiple time scales associated with the dense clusters. While the low and high-frequency limits scale like a simple Maxwell fluid, the crossover cannot be captured with a single timescale and so the moduli cannot be fit to a single Maxwell fluid across all frequencies.

\begin{figure*}[tb]
    \centering
    \includegraphics[width=2\columnwidth]{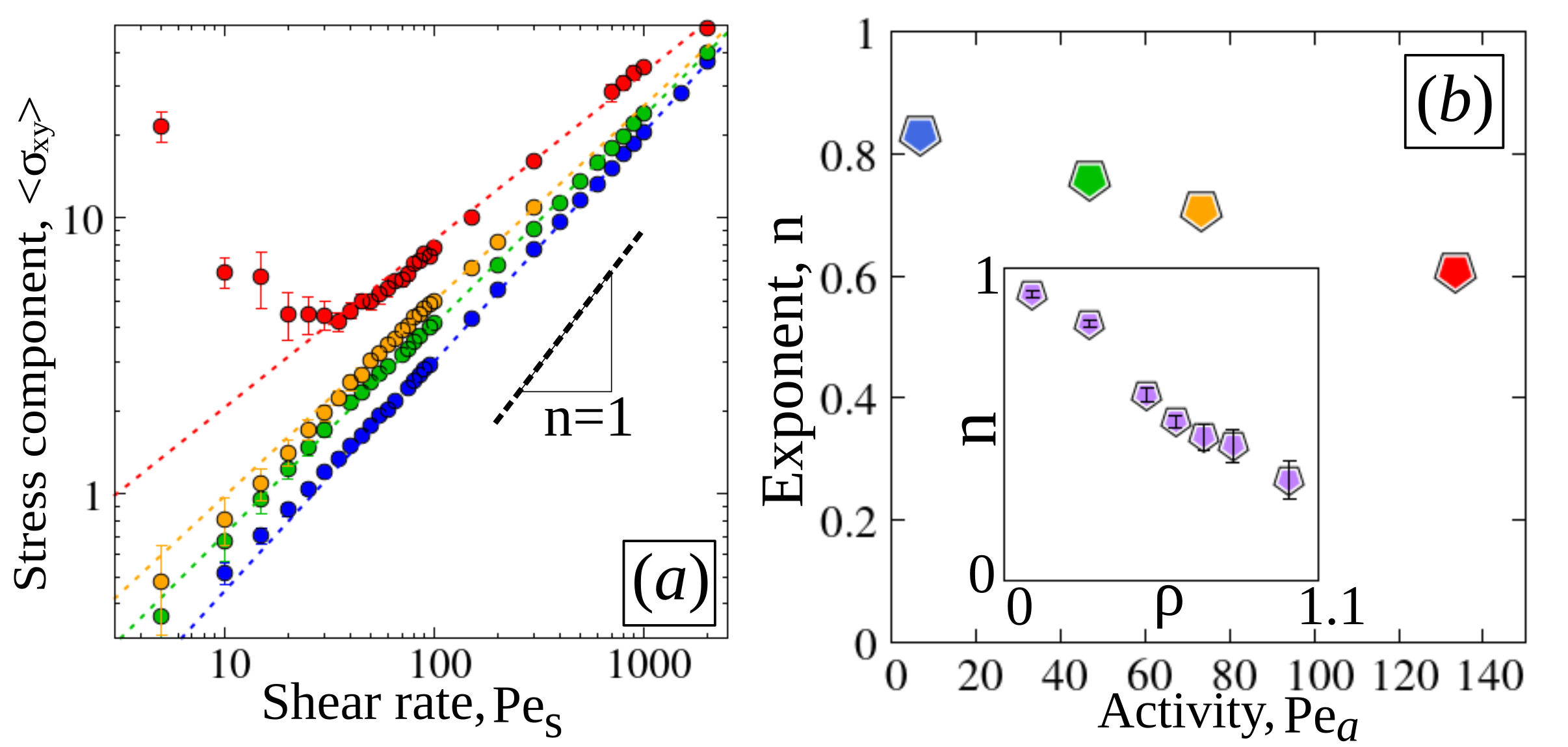}
    \caption{(a) Average out-of-diagonal stress component $\langle \sigma_{xy}\rangle$ for constant shearing as a function of dimensionless shear rate $\pes$ for density $\rho=0.5$ and dimensionless activities $\pea$=6 (blue), 42 (green), 66 (orange) and 120 (red). (b) Power law exponents for the curves shown in panel (a) as a function of dimensionless activity $\pea$ for $\rho=0.5$. 
    \textbf{Inset:} The dependency of the exponent $n$ on density $\rho$ for $\pea=120$.}
    \label{fig:fig4}
\end{figure*}

Overall, the behaviors of the $\gp/\gpp$ (\fig{fig:fig1}) and $\pes^\dagger$ (\fig{fig:fig2}) suggest that suspensions of ABPs both below and above the MIPS transition have a  response akin to a simple Maxwell fluid. 
A Maxwell fluid is characterized by a constant complex viscosity at low frequencies and shear thinning at high frequencies. To  explore this analogy, we  extract the effective shear viscosity using
\begin{equation}
    \eta = \sqrt{\left(\frac{\gp}{\omega}\right)^2+\left(\frac{\gpp}{\omega}\right)^2} \correctMath{.}{,} \label{eq:visc}
\end{equation}
which has been previously employed to uncover non-trivial behavior in active systems~\cite{saintillan2010dilute,heidenreich2011nonlinear,jara2021self}. 

The zero-frequency limit of the viscosity does not increase above passive values (\fig{fig:fig3}; blue and green curves){.} A clear transition from Newtonian  to shear-thinning behavior is observed as $\pes$ increases.  The zero-frequency limit of the viscosity is constant until the MIPS transition, at which point it suddenly begins to rise (\fig{fig:fig3}; inset).  {\color{black} Constant viscosity in the dilute phase is instead explained by the Maxwell relation, $\eta_0 = G'_\infty / \omega_c$. Our data show that both the $G'_\infty$ and the characteristic frequency $\omega_c$ increase with activity in the same way (see SI \cite{SI}). As a result, for small activities their ratio remains roughly constant, which directly accounts for the observed insensitivity of $\eta_0$ below the MIPS transition.}  This constant viscosity in the homogeneous limit is also reflected in \fig{fig:fig1} (blue and green points) where the slope of $G' (\dot{\gamma} \to 0) \sim \eta_0 \, \omega$. {\color{black} Even below the onset of MIPS, activity induces transient clustering that increases local crowding and network connectivity \cite{evans2024re}, potentially leading to a modest rise in the viscosity and a more significant increase in $G'_\infty$, which in turn shifts the crossover Peclet number to higher values.} For this low activity limit, additional activity broadens the range over which the suspension maintains constant viscosity (\fig{fig:fig3}; green). 
The situation is different above the MIPS transition (\fig{fig:fig3}; red), where the zero-frequency viscosity is significantly larger than that of the passive system,  suggesting that  MIPS clusters enhance viscous dissipation. Furthermore, the shear thinning onset above the MIPS transition occurs at frequencies that are smaller than those below the MIPS transition, effectively reducing the Newtonian range. 

At sufficiently large shear ($\pes$), all curves exhibit thinning with the viscosity decaying as $\eta\sim \omega^{-1}$. The shear thinning is caused by particles sliding in layers, similar to homogeneous passive and active colloidal suspensions~\cite{bayram2023motility, wu2009melting}.

To further investigate the high-frequency shear-thinning behavior, we use non-linear constant shear measurements. Under constant shear at high rates, the shear response excites higher order modes. Beyond the crossover, the stress follows a power law~\cite{saramito2016complex} $\sigma_{xy} = B  \dot{\gamma}^n$ (\fig{fig:fig4}a). 
Exponents smaller than $1$ indicate shear-thinning, while $n>1$ is associated to shear thickening and $n=1$ corresponds to a classical Newtonian fluid. \FIG{fig:fig4}b shows the  exponent $n$ for different values of the active \peclet number at a fixed density $\rho=0.5$. 
The inset shows $n$ at a fixed active force ($\pea=120$) as a function of density (see SI~\cite{SI}).
Crucially, the power  decreases monotonically with both parameters and is systematically less than 1, indicating shear-thinning behavior at high frequencies for all activities and densities. 
{ In \fig{fig:fig4}a, the stress for Pe$_a=120$ (red dots) follows an overall non-monotonic trend, which is unlike the simple power law behavior observed below MIPS that is only recovered in the high strain rate limit. This occurs because the dense MIPS structure dominates the rheology of the system at low strain rates, despite the application of a constant strain (see SI~\cite{SI}).  However, when the strain rate is sufficiently large, the structure disappears and the particles follow the direction of the deformation (see SI~\cite{SI}), producing shear thinning ($n<1$) as shown in \fig{fig:fig4}b.} 

Previous studies on the rheology of active colloidal particles under constant shear have reported shear-thickening behavior~\cite{bayram2023motility, wiese2023fluid}.
However, these found shear thickening only for high shear rates and extreme densities~\cite{bayram2023motility}. 
The primary differences between our results and previous studies is their focus on soft overlapping particles at densities well above that of closed-packed disks, while we consider the effect of shear forces on pseudo-hard ABP across the MIPS transition. 
While previous studies have considered the phase behavior of MIPS and proposed useful analogies to coexistance in passive liquid-liquid phase separation~\cite{omar2023}, this study considered the rheological response of MIPS for the first time and revealed a rheological analogy: MIPS responds like a Maxwell fluid  at low and high frequencies, but has an extended crossover regime at intermediate frequencies and a non-monotonic crossover point that follows the active timescale below the MIPS transition and the rotational relaxation timescale above. These results contrast with those found for attractive passive systems, where the crossover time is monotonic with temperature, but the trends for the storage modulus and loss modulus are similar. 
{While for low activity, the activity increases ballistic movement similar to a temperature, at higher activities it causes crowding effects.} This highlights how the MIPS transition fundamentally alters the rheological properties of the system. {\color{black} and the active system can't be described with an effective temperature}.

While our model simplifies several aspects of real active suspensions—such as neglecting hydrodynamic interactions, shear-induced torques, working in two dimensions {\color{black} and boundary effects of the shear plates}—it captures key mechanisms relevant to  active matter. Although the experimental realization of 2D sheared active suspensions remains challenging, our results provide testable predictions for {\color{black} bulk} quasi-2D systems such as active colloids or bacteria confined under shear.
Our findings show that even simple realizations of activity can result in complex  elastic responses since active materials possess alternative routes to respond beyond the thermodynamically accessible pathways of purely passive materials.  However the question of how to fully characterize the rheological properties of active systems remains an open and challenging question, and we hope that our study will inspire more work in this direction.

\section*{Acknowledgements}
C.V. acknowledges fundings
IHRC22/00002 and Proyecto PID2022-140407NB-C21 funded by MCIN/AEI /10.13039/501100011033 and FEDER, UE.
A.C. acknowledges financial support from the National Science Foundation under Grant No. DMR-2321925. This research has also received funding (T.N.S. and K.T.) from the European Research Council under the European Union’s Horizon 2020 research and innovation programme (Grant Agreement Nos. 851196 and 101029079). This work was further supported (K.T.) by the Novo Nordisk Foundation grants no. NNF18SA0035142 and NNF23OC0085012. J.M.R. acknowledges financial support from the UCM predoctoral contract (call CT15/23).
We acknowledge useful discussions with Marco Mazza and James Richards.
For the purpose of open access, the author has applied a Creative Commons Attribution (CC BY) licence to any Author Accepted Manuscript version arising from this submission. 
{The authors thank the Nordita Institute, since the idea of the project was originated during the ``Current and Future Themes in Soft and Biological Active Matter'' workshop, and the Lorentz Center's ``Computational Advances in Active Matter''. } 

\section*{Data Availability Statements}
The data are not publicly available. The data are available from the authors upon reasonable request.

\newpage

%

\end{document}